\documentclass[epj,twocolumn]{webofc}
\usepackage[varg]{txfonts}   
\usepackage{graphicx}
\usepackage{natbib}
\usepackage[english]{babel}
\usepackage{hyperref}

\wocname{epj}
\woctitle{Seismology of the Sun and the Distant Stars 2016}
\begin{document}
\title{Asteroseismic Constraints on the Models of Hot B Subdwarfs: \\ Convective Helium-Burning Cores}
%

\author{\firstname{Jan-Torge} \lastname{Schindler}\inst{1}\fnsep\thanks{\email{jtschindler@email.arizona.edu}} \and
        \firstname{Elizabeth M.} \lastname{Green}\inst{1} \and
         \firstname{W. David} \lastname{Arnett}\inst{1}
}

\institute{Steward Observatory, University of Arizona, 933 North Cherry Avenue, Tucson, AZ 85721,USA
          }

\abstract{

Asteroseismology of non-radial pulsations in Hot B Subdwarfs (sdB stars) offers a unique view into the interior of core-helium-burning stars.
Ground-based and space-borne high precision light curves allow for the analysis of pressure and gravity mode pulsations to probe the structure of sdB stars deep into the convective core.
As such asteroseismological analysis provides an excellent opportunity to test our understanding of stellar evolution.

In light of the newest constraints from asteroseismology of sdB and red clump stars, standard approaches of convective mixing in 1D stellar evolution models are called into question.
The problem lies in the current treatment of overshooting and the entrainment at the convective boundary. 
Unfortunately no consistent algorithm of convective mixing exists to solve the problem, introducing uncertainties to the estimates of stellar ages.
Three dimensional simulations of stellar convection show the natural development of an overshooting region and a boundary layer.
In search for a consistent prescription of convection in one dimensional stellar evolution models, guidance from three dimensional simulations and asteroseismological results is indispensable.

}
\maketitle
\section{Hot Subdwarf B stars}
\label{intro}

Subdwarf B (sdB) stars are a class of hot ($\rm{T}_{\rm{eff}} = 20,000{-}40,000\,\rm{K}$) and compact ($\log g = 5.0{-}6.2$) stars with very thin hydrogen envelopes ($M_{\rm{H}} < 0.01\,M_\odot$) \cite{Heber1986, Saffer1994}.  
They form the so-called extreme horizontal branch (EHB) in the Hertzsprung-Russell diagram, where most of them quietly burn helium in their cores for ${\sim} 10^8$ years.

A violent process stripped them of most of their hydrogen envelope before they underwent the helium flash near the tip of the red giant branch (RGB). 
As a consequence they will not climb up the asymptotic giant branch (AGB) once their core helium is exhausted, but rather evolve directly to become white dwarfs.


Although some sdB stars appear to be single and others occur in wide binaries, surveys have concluded that the majority are in close binaries with white dwarf or low-mass main sequence companions \cite{Maxted2001, Napiwotzki2004, Copperwheat2011}. 
A small fraction of the latter exhibit eclipses, offering insight into the orbital parameters and masses of the systems.  
These HW Virginis stars, named after the discovery system, play a crucial role in determining the mass distribution of sdB stars. 
The median mass of sdB stars estimated from these binaries is $0.469\,M_\odot$ \cite{Fontaine2012}.

The formation scenarios for sdB stars have to explain their existence in both single and binary systems as well as introduce a process by which the hydrogen envelope of the progenitor stars can be lost.
Common envelope ejection (CE) or stable Roche lobe overflow (RLOF) during binary evolution result in systems with close and wide separations, respectively, \citep{Han2002}.
Population synthesis models can well explain the occurrence of binary systems and the mass distribution of the primary and secondary components \cite{Han2003}.
To explain the existence of single sdB stars, extreme mass loss on the RGB \cite{DCruz1996, Sweigart1997, MillerBertolami2008} as well as white dwarf mergers \citep{Webbink1984} were suggested. 
However, the masses from the white dwarf merger channel over-predict the median mass of observed sdB stars \citep{Fontaine2012}.
After the discovery of Earth-sized bodies around sdB stars \cite{Geier2009, Silvotti2007, Charpinet2011b, Geier2012}, it seems possible that sub-stellar companions may play a role in shaping single sdB stars.

Our goal in this article is to discuss the latest results of asteroseismology and their implications for stellar evolution of core helium-burning stars. 
Therefore we will continue by introducing sdB asteroseismology in Section\, \ref{asteroseismology} and the state-of-the-art of sdB stellar evolution in Section\, \ref{stellarevolution}.
In Section\,\ref{convection} we focus on our main discussion of convection in helium-burning cores in light of asteroseismological results.
We present a conclusion to the discussion in Section\,\ref{conclusion}.
For a full review of the field of Hot Subdwarfs we refer the interested reader to the publications of Heber \cite{Heber2009,Heber2016}.

%

\subsection{sdB Asteroseismology}
\label{asteroseismology}
\begin{figure*}[]
 \centering
 \includegraphics[width=0.7\hsize,clip]{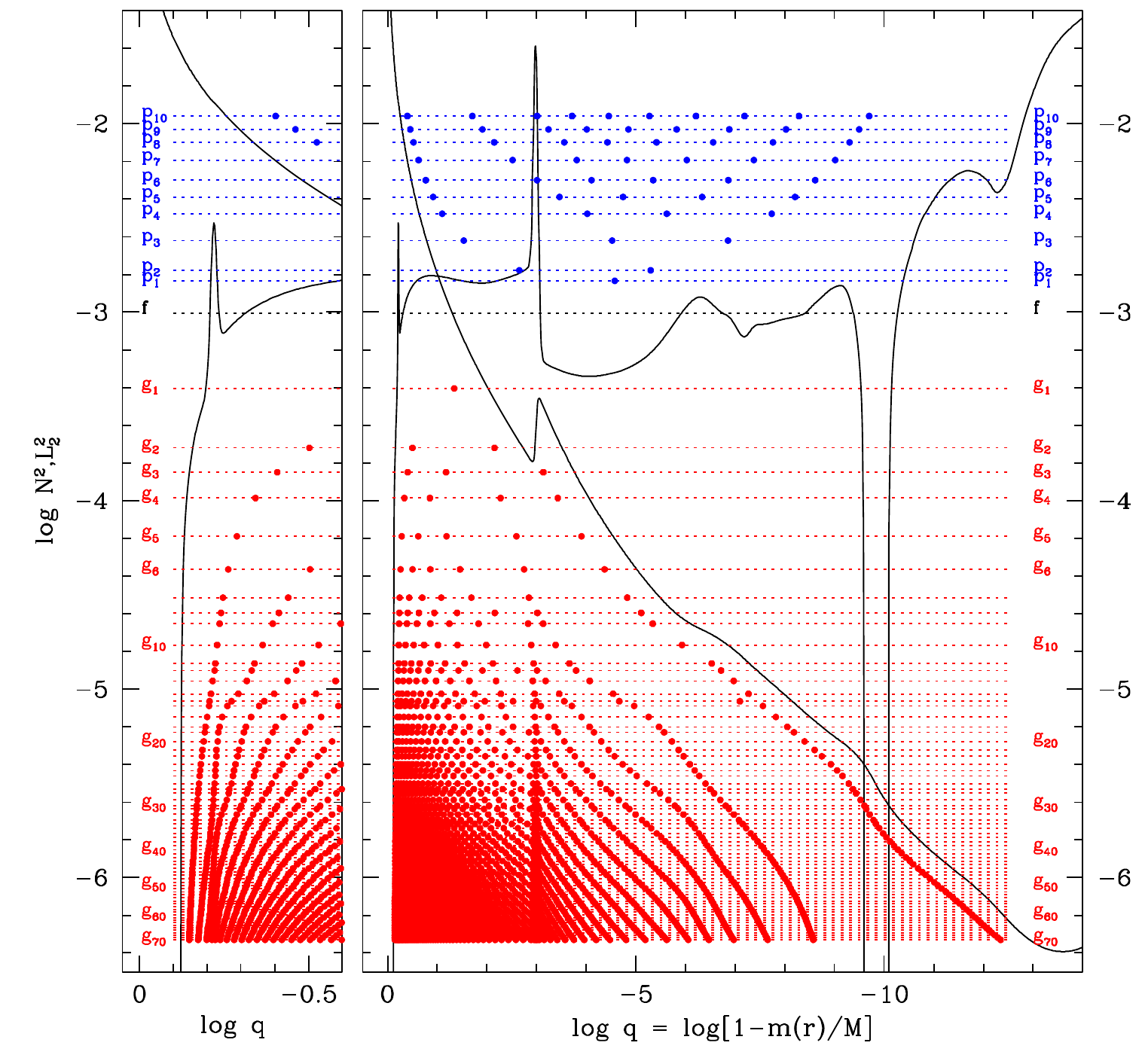}
 \caption{This illustration shows the l=2 modes in a representative sdB model. P-modes propagate where the mode angular frequencies exceed both the Lamb (L2 ) and Brunt-Väisälä (N ) frequencies (blue filled circles show the nodes of modes with order up to k=10). G-modes are confined to the region where the mode angular frequencies are both below the L2 and N frequencies (red filled circles show the nodes of modes with order up to k=70). The left panel shows a close up view of the mode structure in the central region of the star. This Figure was adapted with permission from Charpinet et al. 2013 \cite{Charpinet2013}.
 }
 \label{modes}
\end{figure*}

Short period ($80{-}600\,\rm{s}$) and longer period ($2000{-}14000\,\rm{s}$) stellar pulsations are found in a significant fraction of the sdB star population.
The shorter pressure (p)-mode pulsations were predicted by S.\ Charpinet et al.\ \cite{Charpinet1996} and independently discovered by Kilkenny et al.\ \cite{Kilkenny1997} at nearly the same time. 
Currently more than $60$ p-mode pulsators, termed V361 Hya stars, have been found among the hotter sdB stars. 
These short pulsations are due to low-order, low-degree acoustic waves.

The longer pulsations have their origin in mid to high order, low degree gravity waves (g-modes \cite{Green2003}).
There are about 50 known g-mode pulsators, or V1093 Her stars, on the cooler end of the sdB distribution.
Furthermore, at the intersection of these two populations on the EHB exists a narrow range in which hybrid pulsators can be found  \cite{Schuh2006}, showing both p- and g-mode oscillations.

The non-radial pulsations in sdB stars are opacity driven ($\kappa$-mechanism) \cite{Charpinet1997,Fontaine2003} by partial ionization of iron group elements in their stellar envelopes.
The balance between gravitational settling and radiative levitation creates a region with an overabundance of these iron group elements (especially Fe and Ni, \cite{Jeffery2006}) in the envelopes of these stars, leading to an opacity bump. 
The inclusion of both diffusion processes is therefore not only important to explain the low atmospheric helium abundances, but also to create the driving region for the pulsations.
Since the pulsation regions are by no means pure,
inefficient stellar winds have been proposed to destroy the diffusive balance in the driving region \cite{Chayer2004}.

As can be seen in Figure\,\ref{modes}, p-modes can only propagate in the outer part of the star ($\log q < 0.4$). 
They are reflected back to the surface before reaching the convective core.
G-modes on the other hand can penetrate into the deep interiors ($\log q \lesssim 0.1$). 
The steep chemical transitions between the hydrogen-rich envelope and the helium mantle and between the base of the mantle and the convective C-O-He core result in spikes in the Brunt-Vais\"al\"a-frequency (see Figure\,\ref{modes}). Pulsational modes having nodes (filled circles in Figure\,\ref{modes}) close to these transition regions are either partially trapped above or confined to the lower part of the star. 
This leaves clear signatures on the period spacing of those modes.
The p-mode oscillations are therefore sensitive to the transition between the helium mantle and hydrogen envelope, whereas g-modes also probe the transition of the He mantle and the convective core.
Thus asteroseismological analysis of sdB g-mode pulsators provides an exclusive window into the interior structure of core helium-burning stars.

Quantitative asteroseismological analyses of pulsating sdB stars have been carried out using a forward modeling method \cite{Brassard2001, Charpinet2000, Charpinet2002, Charpinet2002b}. 
The technique uses a neural network of structural sdB models to create theoretical oscillation spectra, which are then compared to the observed frequencies of a given star. 
The best possible match in the given parameter space holds information about the structural parameters of the star. 
Pulsational frequencies derived from light curves of p- and g-mode pulsators using ground-based as well as space-borne (e.g.\,\textit{CoRoT} and \textit{Kepler}) instruments allowed for detailed asteroseismological analysis. 
Also, measurements of the g-mode period spacing in \textit{Kepler} light curves of sdB stars \cite{Reed2011,Ostensen2014} enable comparisons with pulsation properties of stellar evolution models (e.g. see  \cite{Constantino2015}).

Using the forward modeling method \cite[e.g.][]{VanGrootel2008a,Charpinet2008} stellar masses, envelope masses, surface gravities and effective temperatures were constrained and agree remarkably well with measurements from other techniques such as light curve modeling of eclipsing binary systems and spectroscopic analyses (\cite{Green2011} Table 3, \cite{VanGrootel2013}).
As pure structural models, these results are independent of the uncertainties in stellar evolution physics (e.g. convection, nuclear reaction rates or wind mass loss).
Hence, a comparison between stellar evolution models and these asteroseismic measurements may help to identify deficiencies in our current understanding.

%
%

\subsection{Stellar Evolution of Subdwarf B stars}
\label{stellarevolution}
\begin{figure}[h]
 \centering
 \includegraphics[width=\hsize,clip]{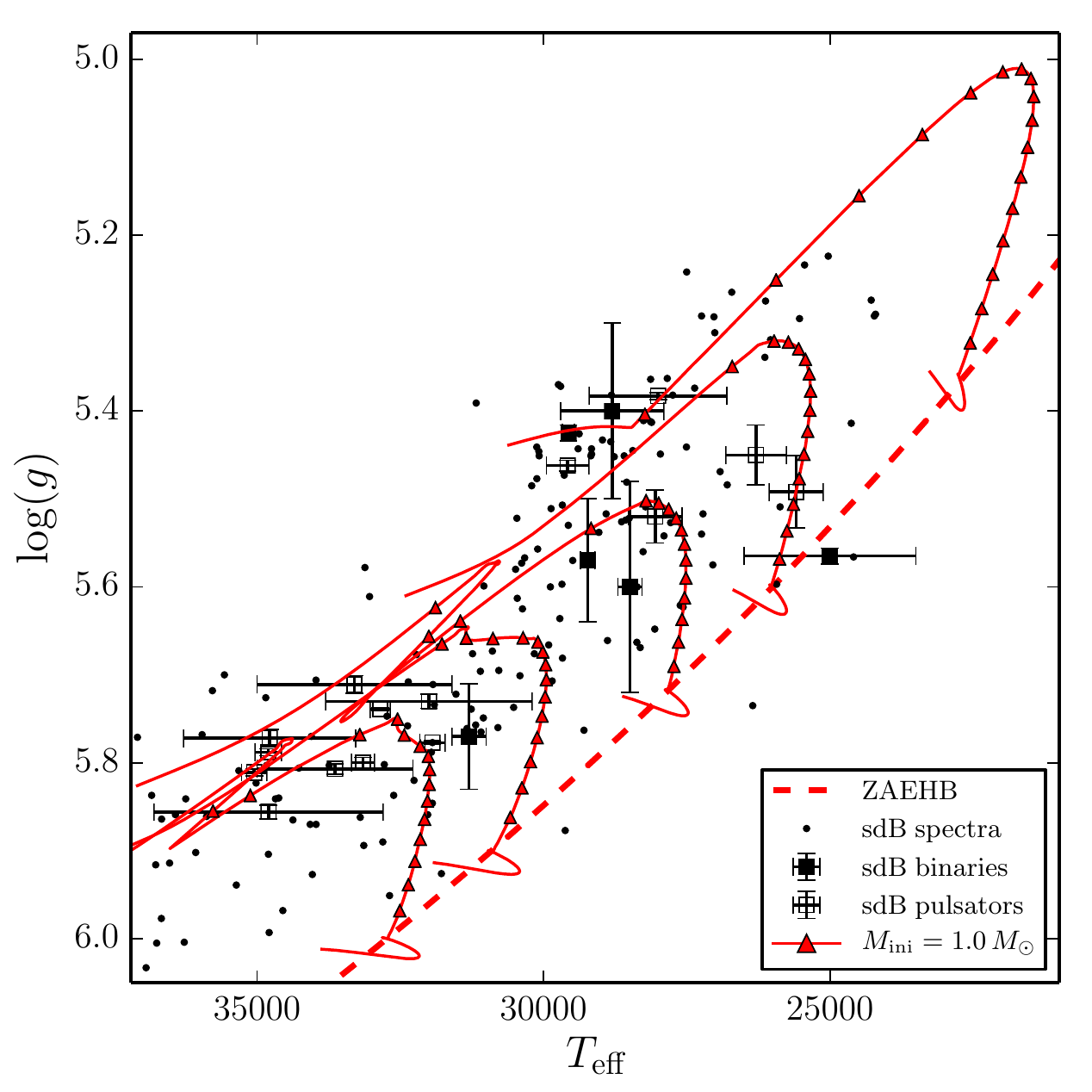}
 \caption{ An example of the evolutionary tracks of sdB stars (solid red curves) in the $\log g{-}T_{\rm{eff}}$ diagram (reproduced from \cite{Schindler2015}). 
 The evolutionary tracks have a progenitor mass of $M_{\rm{ini}} = 1.0\,M_\odot$ and sdB masses of $M_{\rm{sdB}} = 0.4652, 0.4654, 0.4658, 0.4669,
0.4705\,M_\odot$ from bottom to top. The red triangles intersect the lines in intervals of $10^7$ years. 
The dashed line shows the zero age extreme horizontal branch (ZAEHB) for our evolutionary models of $M_{\rm{ini}} = 1.0\,M_\odot$.
The spectroscopic data points (small black dots) for sdB stars \cite{Green2008} agree very well with the open and filled squares with error bars derived from eclipsing binary and asteroseismology analyses, respectively \cite{Fontaine2012}. }
 \label{fig-logteff}
\end{figure}

After stars evolve up the red giant branch and start core helium-burning, losing a substantial amount of their hydrogen envelope in the process, they become horizontal branch stars.
Subdwarf B stars on the extreme horizontal branch have hydrogen envelope masses too low to sustain hydrogen shell-burning.
It is unsurprising that the first systematic stellar evolution studies relevant to sdB stars targeted horizontal branch stars \cite[e.g.][]{Sweigart1976, Sweigart1987, Dorman1993a}.

The detailed physics of the envelope mass stripping due to Roche lobe overflow or common envelopes are mostly neglected in one dimensional stellar evolution models. These processes occur on dynamical timescales of ${\sim}10^3$ years and are usually mimicked by extreme mass loss by stellar winds on the RGB, until only a tiny hydrogen envelope $M_{\rm{env}}\lesssim 0.01\,M_\odot$ remains.
Very recently, physical arguments for common envelope evolution have been introduced to regulate the extreme mass wind loss in \cite{Xiong2016}.

Similarly, older stellar evolution calculations could not handle the violent helium core flash that occurs in $M \lesssim 2.2\,M_\odot$ stars.
Stellar structures at the RGB tip were modified to restart the evolution after the helium flash. 
The He-core material was artificially enriched by $3\%{-}7\%$ of carbon over a region believed to be plausible for the convective mixing during the He-core-flash (see also \cite{Sweigart1997}).
The most widely  adopted models of sdB stars were created using this technique \cite[e.g.][]{Dorman1993a, Charpinet2000, Han2002} and show only slight differences from more recent models which follow the evolution through the He-flash \cite{Serenelli2005}.
Stellar evolution models of sdB stars \cite{Dorman1993a, Han2002} have been successful in explaining the general distribution of the population in the $\log g{-}T_{\rm{eff}}$-diagram and their atmospheric parameters. With the inclusion of physical diffusion processes later studies were further able to self-consistently predict the instability strips for sdB stars \cite{Bloemen2014}. 
However, tension exists between recent asteroseismological analyses of sdB stellar structures and stellar evolution modeling \cite{Constantino2015, Schindler2015}. 
As this is also the main focus of this short review, we will expand this discussion in Section\,\ref{convection}.
In the following paragraphs we highlight some important aspects of sdB stellar evolution models in more depth.

\subsubsection{The He-flash}

Modern state-of-the-art stellar evolution codes are capable of evolving stars from the pre-main-sequence through the Helium flash to the ZAEHB \cite[e.g.][]{Serenelli2005, Paxton2011}. 

However,  two- and three-dimensional simulations of the He-flash \cite{Mocak2008,Mocak2009} have shown that the extent of the convective region during the He-flash, as predicted by standard 1D stellar models, is incorrect.
Turbulent entrainment on both edges of the convection zone lead to rapid growth of the convective region on a dynamic timescale. 
As a consequence the main core flash might not be followed by subsequent mini-flashes leading up to core helium-burning. 

In the current standard algorithms of convection using Mixing Length Theory (MLT, \cite{BohmVitense1958}) turbulent entrainment is not included. 
The convective region in the standard MLT picture only extends throughout the super-adiabatic region, whereas the convective flows in three dimensional simulations naturally extend into sub-adiabatic regions, where the plumes decelerate (``overshooting'').
Additions to the standard algorithms that mimic the physical ``overshooting'' use prescriptions with a free tunable parameter.
Therefore one has to take great caution in interpreting the details of the He-flash calculated by one dimensional stellar models.

\subsubsection{The Importance of Physical Diffusion Processes}

To explain the p-mode pulsations in sdB stars, Charpinet et al. \cite{Charpinet1996} proposed that radiative levitation creates an overabundance of iron group elements in the sdB envelope.
Partial ionization of these metals can then lead to the driving of pulsation via the $\kappa$-mechanism. 
This was confirmed using static models of sdB stars in diffusive equilibrium \cite{Charpinet1997}.
The approach of using static models for the instability calculations was later validated \cite{Fontaine2006}, showing that diffusive timescales are short compared to the stellar evolution. 

In order to explain the g-mode pulsations, Fontaine et al. \cite{Fontaine2003} argued that the same $\kappa$-mechanism was at play.
However, the models exhibiting g-mode pulsations were several thousand Kelvin too cool compared to the distribution of observed pulsators.
This mismatch was termed the blue-edge problem.
The discrepancy could be reduced by using artificially enhanced Fe and Ni abundances in the envelope \cite{Jeffery2006} and the OP \cite{Badnell2005} instead of the OPAL \cite{Iglesias1996} opacities.
In a sequence of pioneering studies, Hu et al. \cite{Hu2009,Hu2010,Hu2011} solved the diffusion equations for gravitational settling, thermal diffusion, concentration diffusion and radiative levitation in stellar evolution models of sdB stars. 
These diffusive processes naturally produce the Fe and Ni enhancements. 
Bloemen et al. \cite{Bloemen2014} finally solved the blue-edge problem by calculating the pulsation properties for stellar models that included the necessary diffusion physics \cite{Hu2011}.

The inclusion of these diffusive processes is therefore crucial for any study focusing on the atmospheric abundances or pulsation properties of sdB stars.

\section{Asteroseismology and Convection in Helium-burning Cores}
\label{convection}

\subsection{A Short Introduction to Convective Mixing in One Dimensional Stellar Evolution}

Most stellar evolution codes treat mixing of convective flows as a ``diffusive'' process. 
A ``diffusion'' operator is chosen for mathematical convenience \cite{Eggleton1972}, based on mixing length theory (MLT, \cite[e.g.][]{BohmVitense1958,Cox1968}).

The extent of the dynamically unstable regions is often assessed by the Schwarzschild or Ledoux criteria for instability.
The Ledoux criterion for convection reads
\begin{equation}
 \nabla_{\rm{rad}} > \nabla_{\rm{ad}} + \frac{\Phi}{\delta} \nabla_{\mu}\ ,
\end{equation}
where
\begin{align}
 \nabla_{\rm{rad}} &\equiv \left( \frac{\partial \ln T}{\partial \ln p} \right)_{\rm{rad}},  &  \nabla_{\rm{ad}} &\equiv \left( \frac{\partial \ln T}{\partial \ln p} \right)_{\rm{ad}},   &  \nabla_{\mu} &\equiv \left( \frac{\partial \ln \mu}{\partial \ln p} \right) 
\end{align}
and
\begin{align}
 \Phi &\equiv \left. \left( \frac{\partial \ln \rho}{\partial \ln \mu} \right) \right|_{P,\mu},  &  \delta &\equiv - \left. \left( \frac{\partial \ln \rho}{\partial \ln T} \right) \right|_{P,T} \ .
\end{align}

If the composition term $\Phi/\delta\ \nabla_{\mu}$ is omitted, one obtains the Schwarzschild criterion for convective instability.
Using the Schwarzschild criterion one refers to the unstable region as super-adiabatic ($\nabla_{\rm{rad}} > \nabla_{\rm{ad}}$) and to the stable region as sub-adiabatic ($ \nabla_{\rm{rad}} < \nabla_{\rm{ad}}$).

When the Ledoux criterion is used, composition gradients are able to stabilize regions that would be unstable to the Schwarzschild criterion.
A receding convection zone during hydrogen core-burning on the main sequence leaves helium enriched material, stable to mixing if the Ledoux criterion was applied, but unstable to the Schwarzschild criterion.
However, the situation is more complicated and double diffusive processes lead to mixing in the seemingly stable region.
This is famously called \textit{semiconvective} mixing and is a needed addition to the standard formulation of MLT, whenever the Ledoux criterion is used \cite[e.g.][]{Langer1985}.

Another addition to canonical convection is \textit{overshoot} mixing.
It refers to the transport of energy and material across the boundary from the dynamically unstable into the stable region.
Since MLT is a local theory, the deceleration and turning of the convective flow is not captured and the convective zone only extends to the edge of the super-adiabatic region. To remedy this, overshooting algorithms \cite[e.g.][]{Zahn1991, Herwig2000} are used to extend the convection region beyond the stable region defined by the instability criterion.

\subsection{The Physics of Convection during Core-helium-burning}

After the helium flash has lifted the degeneracy of the He-core, core-helium-burning starts.
First the triple-alpha process fuses helium into carbon and once a significant carbon abundance has been reached the $^{12}\rm{C}(\alpha,\gamma)^{16}\rm{O}$ reaction dominates the energy generation.
With the start of core-helium-burning a core convection zone develops.

Three dimensional simulations of convection \cite{Meakin2007,Mocak2009,Arnett2009,Viallet2013,Arnett2015} show how plumes of hotter material are accelerated by buoyancy forces in the super-adiabatic region and start to rise up.
Once they reach the sub-adiabatic region, buoyancy braking decelerates the plumes until the flow turns and the material descends back down.
The plumes transport kinetic energy and thermal energy upwards and mix the convective region.
The flow is highly turbulent and the kinetic energy of the flow is converted into internal energy by the physics of the Kolmogorov cascade.

A thin boundary layer develops at the top of the core convection zone. When the convective flow turns, its velocity is perpendicular to the radial coordinate. The material outside the convective zone is not moving and Kelvin-Helmholtz instabilities can develop which slowly entrain material from outside the boundary \cite{Arnett2015}.

The nuclear fusion in the core proceeds to produce carbon and oxygen, which have higher free-free opacities than helium \cite{Castellani1971a}.
Therefore the opacity in the convective core slowly rises with the energy production leading to an increase of the super-adiabatic gradient.
If helium rich material is entrained, carbon-rich (and later oxygen-rich) material is mixed outward of the convective zone increasing the opacity in the stable region in the process.
As a result the boundary region becomes super-adiabatic and the convection zone slowly grows.
Secondly, the downward-mixed helium replenishes the nuclear fuel and the phase of core-helium-burning is prolonged.

In one dimensional models, convective instability is determined by the Ledoux or Schwarzschild criterion. 
Mixing length theory (MLT) predicts the convective velocity locally and calculates the convective energy flux for the energy balance.
However, fluxes of the convective velocity across zone boundaries are not included in this framework and therefore deceleration of material (physical overshooting) cannot exist in this picture.
Without the addition of ``overshoot'' prescriptions, an extremely sharp abundance gradient develops at the boundary where convective neutrality ($\nabla_{\rm{rad}} = \nabla_{\rm{ad}}$ for the Schwarzschild criterion) is violated.
This framework does not allow for entrainment of material at the boundary and as a consequence the convective zone is unable to grow.
Historically \cite{Castellani1971a} overshoot prescriptions are used to circumvent this problem. 
However, they usually introduce a free tunable parameter and therefore the stellar evolution model loses its ``predictive'' power.
Alternatively, concentration diffusion has been shown to soften the abundance gradient and naturally lead to a growth of the convection zone \cite{Michaud2007}. 
Yet, the rate of entrainment through diffusion might not necessarily be the same as for convective entrainment.
Both approaches do satisfy convective neutrality at the boundary.

This boundary problem is closely related with another phenomenon.
It has been shown that the super-adiabatic gradient develops a minimum during core-helium-fusion. \cite{Paczynski1970,Castellani1971b,Demarque1972,Robertson1972}. 
Over time the super-adiabatic gradient can decrease until the minimum reaches convective neutrality. 
Since MLT defines convection to occur in regions with super-adiabatic excess, the definition of the convective zone becomes ambiguous at this point. 
In consequence standard algorithms will force the convection zone to split.
As a solution the material beyond the minimum was mixed with physically motivated but ad-hoc schemes to satisfy convective neutrality in that region. 
Many similar dedicated algorithms with different approaches and terminology have been developed to achieve this (e.g.\, ``induced semiconvection'' \cite{Castellani1971b}, ``semiconvection'' \cite{Robertson1972}, ``partial mixing'' \cite{Dorman1993b}, ``maximal overshooting'' \cite{Constantino2015}).
We will refer to this process as ``induced semiconvection'' for the rest of this article.

\begin{figure*}[ht]
 \centering 
 \includegraphics[width=0.95\hsize,clip]{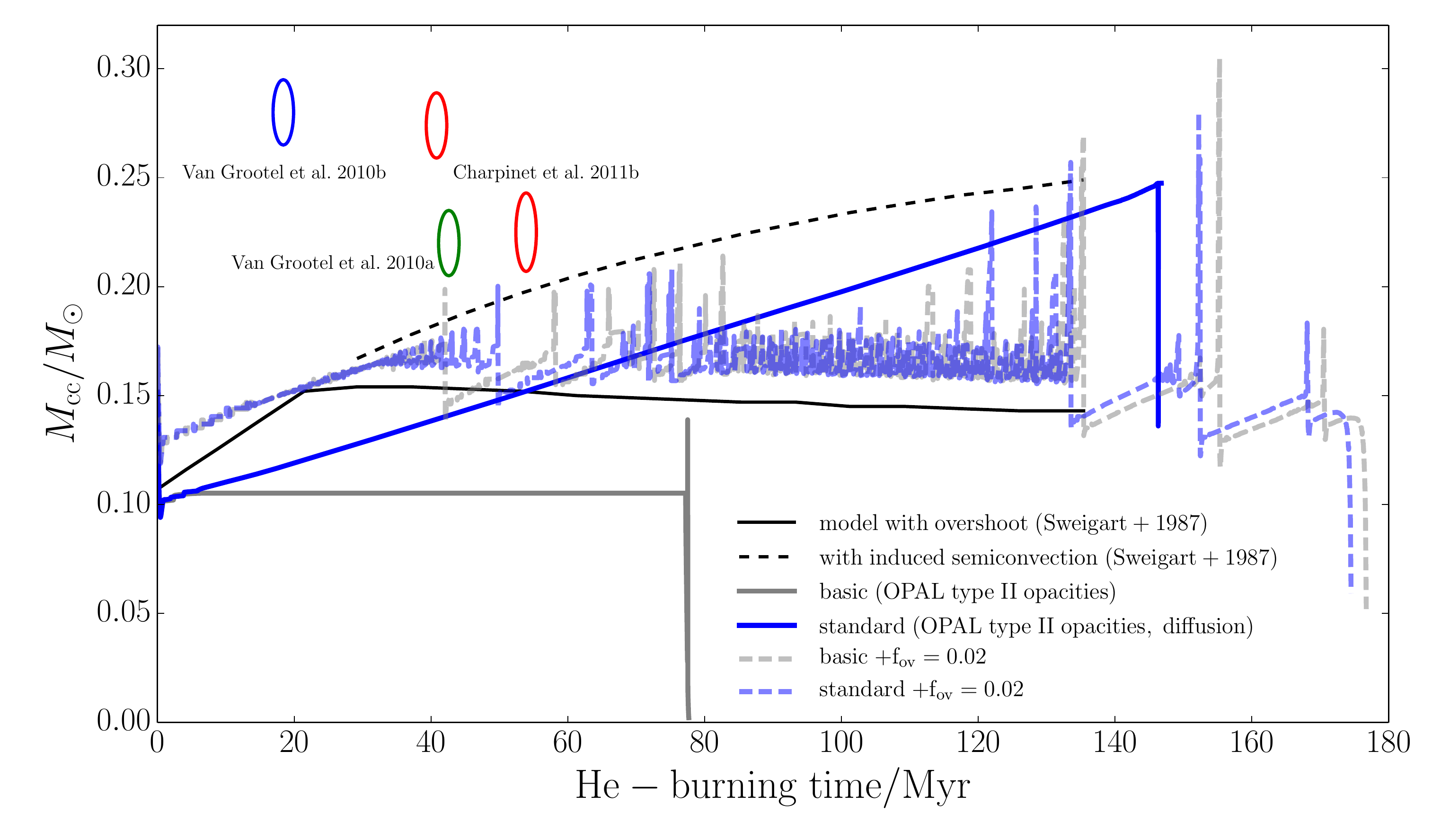}
 \caption{The extent of the convective cores as a function of time for four MESA models \cite{Schindler2015} (gray and blue curves) and two older models from Sweigart et al. \cite{Sweigart1987} (black curves). The Sweigart models include either overshoot or induced semiconvection.  A comparison of the basic and standard MESA models (solid curves) shows the effect of including diffusion in the absence of overshoot.  The last two MESA models are the same, except that overshoot is now included.
The largest cores result from either overshoot or concentration diffusion to induce core growth. Large values of overshoot do not produce core growth, just larger cores. The rapid fluctuations in the models with overshoot suggest that the mixing algorithm experiences numerical instabilities at the boundary. Finally, we contrast the simulations with asteroseismic results (ovals) from the analysis of three g-mode pulsators \cite{VanGrootel2010a,VanGrootel2010b,Charpinet2011}.}
 \label{fig_convcore}
\end{figure*}

However, the convection zone does not necessarily split. In some of our models, which include concentration diffusion to enable core growth, the super-adiabatic minimum never decreases to convective neutrality (e.g. see the model with the solid blue line Figure\,\ref{fig_convcore} \cite{Schindler2015}).

Core convection during helium-burning remains a complex problem of stellar evolution.
We have briefly mentioned the 3D picture of convection and then discussed core convection during core-helium-burning in one dimensional models.
It is time we turned towards observational evidence for guidance and constraints.

\subsection{Tension Between 1D Stellar Evolution Models and Asteroseismology}

The advent of quantitative asteroseismological analyses in the era of satellite-borne precision photometry opens the possibility to study the deep interior of stellar structures.
The results of asteroseismology allow us to contrast stellar evolution models with measurements from real stars and can help to constrain the inadequacies of our current modeling prescriptions.
Here we will focus on results regarding core-helium-burning convection, but asteroseismology also yields measures of the internal rotation.

\subsubsection{Subdwarf B stars}
In contrast with red clump or horizontal branch (HB) stars, sdB stars cannot sustain hydrogen shell-burning nor an outer convective hydrogen envelope. 
This allows for direct observations of g-mode pulsations in core-helium-burning stars, which makes them unique probes of the core convection zone. 
Out of the ${\sim}15$ known g-mode pulsators for which there are sufficient photometric observations for asteroseismology, three have been analyzed using the forward modeling technique described above.

The results directly constrain the extent of the core convection zone as well as the nature of the abundance gradient at the boundary.
The analyses for the three pulsators estimate the convection zone to extend out to $M_{\rm{cc}} = 0.22 \pm 0.01\,M_\odot$ \cite{VanGrootel2010a}, $M_{\rm{cc}} = 0.28 \pm 0.01\,M_\odot$ \cite{VanGrootel2010b} and either $M_{\rm{cc}} = 0.273_{-0.010}^{0.008}\,M_\odot$ or $M_{\rm{cc}} = 0.225_{-0.016}^{+0.011}\,M_\odot$ \cite{Charpinet2011b}.
In the last case two equally probable solutions were found.
The abundance gradients at the convection zone boundary indicate that all three stars are significantly less than halfway through their He-burning lifetimes, having consumed only about 20\%-40\% of the helium in their cores.

We conducted a study to test whether we could reproduce the interior structures inferred from asteroseismology using standard algorithms in one dimensional stellar evolution models \cite{Schindler2015}. 
We carried out these calculations with the Modules for Experiments in Stellar Astrophysics (MESA, \cite{Paxton2011,Paxton2013}).
While our standard model used concentration diffusion to allow for convective core growth (see also \cite{Michaud2007}), it was not able to reproduce the larger core sizes for these three stars.
Only extreme values for additional (exponential) overshoot (as implemented in MESA \cite{Paxton2011}) were able to bring the extent of the convective core in marginal agreement with the asteroseismological results.

We display the growth of the convective cores for a range of model physics in Figure\,\ref{fig_convcore} and contrast them with results of Sweigart et al. \cite{Sweigart1987} using overshoot and induced semiconvection.
Our ``standard'' model shows a monotonically growing convective core that reaches the same extent as the model with induced semiconvection.
The convective boundary seems to fluctuate for our models with overshoot, indicating an unstable behavior that might well be numerical in nature.
However, none of the convective core sizes match the results from asteroseismology, indicated by the colored error ellipses.

Assumptions inherent in the parameterization of the stellar structures in the forward modeling may also introduce systematic effects.
In particular, estimates of the abundance gradient are very sensitive to its assumed shape. 
While this could introduce uncertainties in the core He-burning lifetime of those three stars, the standard algorithms of convection still seem inadequate to reproduce the inferred stellar structures.

\subsubsection{Horizontal Branch and AGB Stars}
The period spacing of g-modes is very sensitive to the boundary of the convective helium-burning core, as we discussed above (Section\,\ref{asteroseismology}).
In contrast to sdB stars, red clump stars are \cite{Fontaine2012} core He-burning stars which retained their hydrogen envelopes. 
Light curves of these stars measured by the NASA \textit{Kepler} mission allowed the g-mode spacing to be inferred from mixed modes for hundreds of these stars \cite{Mosser2012,Mosser2014,Pinsonneault2014}.

Bossini et al. \cite{Bossini2015} constructed stellar evolution models using multiple stellar evolution codes and studied the influence of different convective prescriptions in comparison to the observational data.
The goal was to find the convection prescription that would match the AGB luminosity at the AGB bump as well as describe the measured period spacing.
While standard models (including induced semiconvection and overshoot) would reproduce the AGB bump luminosity, they could not describe the period spacing. However, models with extreme overshoot reproduced the period spacing but failed to match the AGB bump luminosity.
The authors then outlined a candidate model with a moderate overshooting region characterized by adiabatic stratification to fit both constraints, for use in further studies. 

Another study conducted by Constantino et al. \cite{Constantino2015, Constantino2016} tested how well a range of stellar models with different convection prescriptions could reproduce the g-mode period spacing as well as the relative numbers of stars on the AGB and RGB in clusters.
They demonstrated that the period spacing could only be matched using their new ``maximal overshoot'' mixing scheme, which produces the largest convective cores possible. Yet, they also stressed that mode trapping can bias the observationally inferred value of the period spacing. If so, the standard model using induced semiconvection would also suffice.
Both solutions could explain the cluster counts, if mixing beneath the Schwarzschild boundary during subsequent early-AGB evolution occurred in the ``maximal overshoot'' models or if models with induced semiconvection could avoid helium core breathing pulses.

Neither study provides a clear way forward for the use of convection prescriptions during this phase of stellar evolution. 
But they do identify important constraints for one dimensional stellar evolution models.

\subsubsection{White Dwarfs}

Ongoing asteroseismic analyses of the chemical abundance structure of white dwarfs with carbon-oxygen cores (CO-WD) might deliver further clues to understand convection during core-helium-burning (N. Giammichele, private communication). 
The oxygen abundance profile of a CO-WD is set by the interplay of core-helium-burning and the convection zone as well as the outward moving helium-burning shell and thermal pulses on the AGB.
Estimates of the abundance profile in comparison with 1D stellar evolution calculations could provide new constraints on commonly used convection prescriptions.

\subsection{On The Issue of Core Breathing Pulses}

An interesting phenomenon encountered in low mass stars at the end of the helium-burning lifetime, when the helium abundance is very low, are the so-called ``breathing pulses'' \cite{Sweigart1973, Castellani1985}.
During ``breathing pulses'' the convective zone grows rapidly, mixes new helium fuel down into the core and collapses again.
In most cases this effect revives the core and prolongs the helium-burning lifetime considerably.
In the late stage of core-helium-burning, the reaction rates depend strongly on the helium abundance. 
Mixing convective boundary layers with a sharp abundance gradient will lead to a sudden increase in the energy generation rate and can explain the fast expansion of the convective core.
Whether this situation occurs in ``real'' stars is doubtful.
Dorman et al. \cite{Dorman1993b} argued that this behavior is exclusively numerical in nature.
In the more recent study by Constantino et al. \cite{Constantino2016}, breathing pulses are found in stellar models with overshoot which exhibit a sharp abundance gradient at the convective boundary.
The authors show that a monotonically decreasing helium abundance in models with partial mixing, induced semiconvection or maximal overshoot ensures the stability of the convective core.
In fact, they point out that non-local treatments of overshooting (see also \cite{Bressan1986}) seem to avoid avoid breathing pulses naturally \cite{Meakin2007}.

A variety of studies comparing observational constraints to theoretical models \cite{Bressan1986, Caputo1989, Cassisi2001, Constantino2016} have led to the conclusion that breathing pulses do not occur in natural environments.

It is the nature of the helium abundance gradient and the rate of entrainment at the convective boundary during core 
helium-burning that allow for breathing pulses to occur.
Their non-existence provides a strong constraint on the prescriptions of convective mixing and the treatment of the convective boundary.



\subsection{The $^{12}\rm{C}(\alpha,\gamma)^{16}\rm{O}$ Nuclear Rate and Convection}

The newest measurements of the $^{12}\rm{C}(\alpha,\gamma)^{16}\rm{O}$ nuclear reaction rate (NACRE II, \cite{Xu2013}) have reduced the uncertainty on its value. 
However, in order to reach the conditions for this reaction in low mass stars, the measured rate has to be extrapolated to lower energies, thereby introducing new uncertainties. 

Constantino et al. \cite{Constantino2015} considered the effect of the $^{12}\rm{C}(\alpha,\gamma)^{16}\rm{O}$ nuclear reaction rate on the g-mode period spacing, and found that doubling the rate coefficient did not help the stellar models to achieve the larger period spacing observed in red clump stars. 
Yet, they noted that a higher $^{12}\rm{C}(\alpha,\gamma)^{16}\rm{O}$ rate increased the convective core mass.

Asteroseismological studies of white dwarfs \cite{Metcalfe2002, Metcalfe2003} inferred the central oxygen abundance to draw conclusions on the effective $^{12}\rm{C}(\alpha,\gamma)^{16}\rm{O}$ nuclear reaction rate over the star's lifetime. 
However, great care has to be taken since diffusion \cite{Fontaine2002} or convection can alter the stellar model's pulsation spectrum and may lead to wrong conclusions.

In fact, core convection during core-helium-burning and the $^{12}\rm{C}(\alpha,\gamma)^{16}\rm{O}$ reaction feature a complex interplay \cite{Arnett1996}.
The growth of the convective core depends on the entrainment of more opaque material into the stable region to incite instability. 
The increase in opacity in turn is due to the carbon and oxygen yields of the reactions of helium-burning.
So a higher $^{12}\rm{C}(\alpha,\gamma)^{16}\rm{O}$ nuclear rate leads to a higher energy production rate  and a larger oxygen yield.
Both of these lead to a growing convective core if entrainment or some sort of boundary mixing is included.

On the other hand, the rate of entrainment determines the amount of helium brought into the burning region and therefore influences the rate of the $^{12}\rm{C}(\alpha,\gamma)^{16}\rm{O}$ reaction as well.
It can be easily shown that the central oxygen abundance depends on the entrainment rate or on the specified amount of ``overshoot'' at the convective boundary.

These effects complicate studies that like to infer the $^{12}\rm{C}(\alpha,\gamma)^{16}\rm{O}$ nuclear reaction rate from asteroseismology by introducing uncertainties that depend on the adopted algorithm of convective entrainment and vice versa.

\section{Conclusion}
\label{conclusion}

Space-borne precision photometry has stimulated asteroseismological analysis for a variety of stars. 
Regarding sdB stars much progress has been made in the last fifteen years.
It has become clear that the physics of radiative levitation and gravitational settling are needed to reproduce the pulsational instabilities through iron element opacity bumps in their envelopes.
Analyses of g-mode pulsations constrained the extent of the convective core, the nature of the abundance gradient as well as internal rotation profiles.

While the instability strips can now be reproduced, difficulties for coherent stellar evolution models still persist.
In the binary evolution channels, the prospective sdB star loses most of its hydrogen envelope mass during the common envelope or the stable Roche lobe overflow phase. 
Current stellar evolution models utilize some form of extreme mass loss prescription to mimic this dynamical event.
After the star is stripped of most of its hydrogen envelope, it undergoes the He-flash.
Although many recent stellar evolution codes can follow the evolution through this violent event, three dimensional simulations show that the one dimensional picture is not fully correct.

Lastly the standard algorithms employed to simulate convection in stellar evolution models seem to be inadequate during the phase of core-helium-burning.
Asteroseismology of g-mode sdB pulsators indicates larger convective cores than can be reproduced in the current framework of mixing algorithms. 
Recent studies compared the g-mode period spacing found in red clump stars to stellar evolution models.
The authors demonstrate that the current, mostly ad-hoc, treatments of overshooting and/or induced semiconvection can reproduce the observations in a few cases.
The difficulties can be traced back to the treatment of physical overshooting and the entrainment at the convective boundary.
Unfortunately no consistent prescription of core-helium-burning convection using the standard algorithms has emerged yet.
This also applies to other stages of stellar evolution.
With recent asteroseismological analyses constraining the convective cores during main-sequence evolution \cite{Deheuvels2016, Moravveji2016}, a consistent picture might well emerge in the coming years.

Where the asteroseismological results constrain the long term behavior of convection on the stellar structure, three dimensional models of convection offer insight into the dynamics of the physical mechanisms in play.
These simulations show the development of a boundary layer and the strong fluctuations of the turbulent convective flow.
Physical overshooting and entrainment occur naturally when the moving convective plumes enter the sub-adiabatic region, are slowed down and turn.

In order to develop a consistent prescription for convection in evolutionary models, one has to undertake the dangerous procedure of integrating over the turbulent fluctuations.
One possibility is to use the 3D simulations to provide closure to the Reynolds averaged Navier-Stokes equations. Approximations on the basis of these equations might allow for implementation in a stellar evolutionary code without calibrations to astronomical data \cite{Arnett2015}.

The recent insight into stellar structures from asteroseismology and convective behavior from 3D simulations has highlighted the known inadequacies in our stellar evolution models, but also provided constraints to tackle them. 
Still, convection, especially in core-helium-burning stars, remains a fundamental problem of modern astrophysics for now.

\subsection*{Acknowledgements}
JTS thanks the SOC of ``Seismology of the Sun and the Distant Stars 2016'' and the Theoretical Astrophysics Program at the University of Arizona for support.

%
%
\bibliographystyle{woc}


\end{document}